\documentclass[10pt,conference]{IEEEtran}
\IEEEoverridecommandlockouts
\usepackage{cite}
\usepackage{amsmath,amssymb,amsfonts}
\usepackage{algorithmic}
\usepackage{graphicx}
\usepackage{textcomp}
\usepackage[dvipsnames]{xcolor}
\usepackage[utf8]{inputenc}
\usepackage[english]{babel}
\usepackage{breakcites} 
\usepackage[printonlyused, nolist]{acronym}
\usepackage{graphicx} 
\usepackage{pgfplots} 
\pgfplotsset{compat=newest}
\usepackage{colortbl}
\usepackage[breakable, hooks]{tcolorbox}
\usepackage{array}
\usepackage{enumitem}
\usepackage{pgf-pie}
\usepackage{hyperref}

\makeatletter

\def\tcb@restore@footnote{%
  \def\@mpfn{footnote}%
  \def\thempfn{\arabic{footnote}}%
  \let\@footnotetext\tcb@footnote@collect
}

\long\def\tcb@footnote@collect#1{%
  \expandafter\gappto\expandafter\tcb@footnote@acc\expandafter{%
    \expandafter\footnotetext\expandafter[\@thefnmark]{#1}%
  }%
}

\def\tcb@footnote@use{%
  \tcb@footnote@acc
  \global\let\tcb@footnote@acc\@empty
}
\global\let\tcb@footnote@acc\@empty

\tcbset{
  every box/.style={
    before upper pre=\tcb@restore@footnote
  },
  every box on layer 1/.append style={
    after app=\tcb@footnote@use
  }
}
\makeatother

\usepackage{vcell}
\usepackage{caption}

\def\BibTeX{{\rm B\kern-.05em{\sc i\kern-.025em b}\kern-.08em
    T\kern-.1667em\lower.7ex\hbox{E}\kern-.125emX}}
\begin{document}

\title{Energy-Saving Strategies for Mobile Web Apps \\and their Measurement:\\ Results from a Decade of Research}

\author{\IEEEauthorblockN{Benedikt Dornauer\IEEEauthorrefmark{1} and Michael Felderer\IEEEauthorrefmark{2}}
\IEEEauthorblockA{\IEEEauthorrefmark{1}\IEEEauthorrefmark{2}\textit{University of Innsbruck}, 6020 Innsbruck, Austria}
\IEEEauthorblockA{\IEEEauthorrefmark{1}\IEEEauthorrefmark{2}\textit{University of Cologne}, 50923 Cologne, Germany}
\IEEEauthorblockA{\IEEEauthorrefmark{2}\textit{German Aerospace Center (DLR), Institute for Software Technology}, 51147 Cologne, Germany}
\IEEEauthorblockA{ORCID: \IEEEauthorrefmark{1}0000-0002-7713-4686, \IEEEauthorrefmark{2}0000-0003-3818-4442}
}

\maketitle

\begin{abstract}
In 2022, over half of the web traffic was accessed through mobile devices. By reducing the energy consumption of mobile web apps, we can not only extend the battery life of our devices, but also make a significant contribution to energy conservation efforts. For example, if we could save only 5\% of the energy used by web apps, we estimate that it would be enough to shut down one of the nuclear reactors in Fukushima. 
This paper presents a comprehensive overview of energy-saving experiments and related approaches for mobile web apps, relevant for researchers and practitioners. 
To achieve this objective, we conducted a systematic literature review and identified 44 primary studies for inclusion.
Through the mapping and analysis of scientific papers, this work contributes: (1) an overview of the energy-draining aspects of mobile web apps, (2) a comprehensive description of the methodology used for the energy-saving experiments, and (3) a categorization and synthesis of various energy-saving approaches. 
\end{abstract}

\begin{acronym}
    \acro{pwa}[PWA]{Progressive Web App}
    \acro{css}[CSS]{Cascading Style Sheets}
    \acro{html}[HTML]{Hypertext Markup Language}
    \acro{js}[JS]{JavaScript}
    \acro{slr}[SLR]{Systematic Literature Review}
    \acro{cpu}[CPU]{Central Processing Unit}
    \acro{led}[LED]{Light Emitting Diode}
    \acro{wifi}[Wi-Fi]{Wireless Fidelity}
    \acro{os}[OS]{Operating System}
    \acro{mcpd}[MCPD]{mobile cross-platform development}
    \acro{iot}[IoT]{Internet of Things}
    \acro{qos}[QoS]{Quality of Service}
    \acro{cwv}[CWV]{Communicating Web Vessels}
    \acro{oled}[OLED]{Organic Light-Emitting Diode}
    \acro{lcd}[LCD]{Liquid Crystal Display}
    \acro{gsm}[GSM]{Global System for Mobile Communications}
\end{acronym}
\renewcommand{\IEEEiedlistdecl}{\relax}
\section{Introduction}

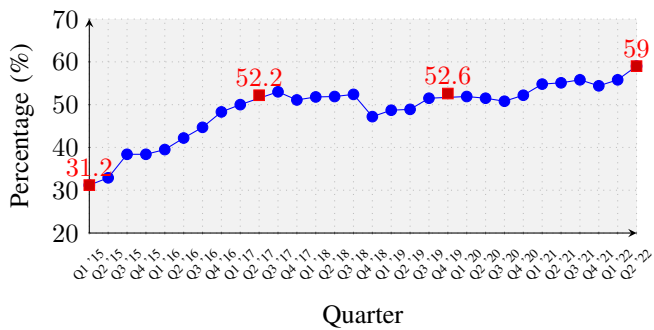
\begin{figure}[htbp]
    \centering
    \begin{tikzpicture}
    \begin{axis}[
    width=\linewidth,
    height=0.5\linewidth,
    axis lines=left,
    major tick length=1,
    ymajorgrids=false,
    xmajorgrids=false,
    xlabel=Quarter,
    ylabel=Percentage (\%),
    ylabel style={yshift=0pt},
    ymin=20, ymax=70, 
    xtick={1,2,...,30}, 
    grid,
    grid style=dotted,
    axis background/.style={fill=gray!10},
    xticklabels={
    Q1 '15,Q2 '15,Q3 '15,Q4 '15,Q1 '16,Q2 '16,Q3 '16,Q4 '16,Q1 '17,Q2 '17,Q3 '17,Q4 '17,Q1 '18,Q2 '18,Q3 '18,Q4 '18,Q1 '19,Q2 '19,Q3 '19,Q4 '19,Q1 '20,Q2 '20,Q3 '20,Q4 '20,Q1 '21,Q2 '21,Q3 '21,Q4 '21,Q1 '22,Q2 '22},
    xticklabel style={font=\tiny},
    ytick={0,10,...,100},
    legend pos=north west,
    xticklabel style={rotate=45}
    ]
    
    \addplot[color=blue,mark=*] coordinates {
    (1, 31.2)
    (2,32.9)
    (3,38.4)
    (4,38.4)
    (5,39.5)
    (6,42.2)
    (7,44.7)
    (8,48.3)
    (9,50.0)
    (11,53.0)
    (12,51.1)
    (13,51.8)
    (14,51.9)
    (15,52.4)
    (16,47.2)
    (17,48.7)
    (18,48.9)
    (19,51.5)
    (21,51.9)
    (22,51.5)
    (23,50.8)
    (24,52.2)
    (25,54.8)
    (26,55.1)
    (27,55.8)
    (28,54.4)
    (29,55.8)
    (30,59.0)
    };

    \addplot+[only marks,nodes near coords] table[meta=label,y index=1,x index=0] {
    x y label
    1 31.2 31.2\%
    10 52.2 52.2\%
    20 52.6 52.6\%
    30 59.0 59.0\%
    };

    \end{axis}
    \end{tikzpicture}
    \caption{Percentage of global mobile website traffic from Q1 2015 to Q2 2022~\cite{internetlivestatsTotalNumberWebsites2022}.}
    \label{fig:mobileWebTraffic}
\end{figure}

The percentage of mobile web traffic has increased over the last seven years,  as shown in Figure \ref{fig:mobileWebTraffic}. This trend is expected to continue, leading companies to invest further into (mobile) web development~\cite{kendeGlobalConnectivityReport2022}.
\\
In this paper, we use the term \textit{mobile web app} to refer to an application that is an optimized website for mobile access (smartphone, tablets) ~\cite{ chan-jong-chuInvestigatingCorrelationPerformance2020, serranoMobileWebApps2013}. In contrast to a native application, which the end user must install on their device, web apps provide widespread compatibility, fast development, and complete application portability ~\cite{leePridePrejudiceProgressive2018, malavoltaWebbasedHybridMobile2016}. This application category also encloses "old-fashioned" responsive mobile websites designed to adjust their layout and content automatically to provide an optimal viewing experience on a wide range of devices. \ac{pwa} is a relatively novel approach that combines the flexibility of responsive web content with the functionality of native apps, such as offline capabilities, push notifications, and home screen installation ~\cite{nunkesserWebNativeHybrid2018}. \ac{pwa} still uses traditional \ac{html}, \ac{css}, and \ac{js}, or available offshoots like Typescript, SASS or Jade. This type of application can also be seen as a kind of \ac{mcpd} because they are built using web technologies that are compatible with a wide range of devices and platforms. Also, other \ac{mcpd} are built on web technologies like React, Ionic, or Fuse ~\cite{majchrzakComprehensiveAnalysisInnovative2017} and count as mobile web apps. Moreover, WebViews, common for displaying web content on a native app, are also some kind of mobile web app ~\cite{BuildWebApps2022}. 

By consuming different types of web content, users' devices are constantly expending varying levels of energy. If we can save energy on our phones, we can extend the battery life and reduce the number of times we need to recharge.  As a result, users can enjoy more extended periods of use between charges and avoid the frustration of a dead battery.

Additionally, with the fast progress of global warming, steps are needed to save energy. In this way, the term Green \ac{iot} has been raised. Green \ac{iot} aims to reduce the environmental impact of IoT devices by minimizing their energy consumption and promoting sustainable practices ~\cite{arshadGreenIoTInvestigation2017}. 

The following example demonstrates the significant impact that energy-saving techniques on smartphones can have. 

\begin{tcolorbox}[breakable,  boxsep = 0pt, outer arc = 4pt]
\vspace{2mm}
\hspace{2.0mm} A data set provided by~\cite{SmartphoneBatteryLife2023} comprised information on the battery life of 150 smartphones from 2022. The data includes information about the battery capacity [mAh]. In addition, it contains for each phone the duration [hh:mm] on how long it is able to process and display web content via \ac{wifi} before fully discharging. 

\hspace{2.5mm}The data revealed that the average capacity of these devices is 4818 mAh, and it takes an average of 13 hours and 3 minutes to fully discharge while consuming web content.
By considering different energy improvement percentages, we derived the following Table \ref{tab:energySaving}. The table shows inter alia the expected improvement in the discharge time of an average smartphone from 2022. 

\captionsetup{type=table}
\begin{center}
    \captionof{table}{Impact of energy-savings.}
    \label{tab:energySaving}
    {\footnotesize
    \begin{tabular}
    {m{3.0cm}m{0.5cm}m{0.6cm}m{0.7cm}m{0.7cm}} 
   \hline
\textbf{Battery-savings [\%]} & \textbf{5\%} & \textbf{10\%} & \textbf{20\%} & \textbf{30\%} \\ 
\hline
Increase of web browsing duration before battery depletion after a full charge \textbf{[hh:mm]:} & 0:39 & 1:18 & 2:36 & 3:55 \\ 
\hline
Battery capacity savings after full charge \textbf{[mAh]:} & 241 & 482 & 964 & 1446 \\ 
\hline
The average daily battery capacity savings for mobile web users \textbf{[mAh]}: & 68 & 136 & 272 & 408 \\ 
\hline
Total battery power of all mobile users \textbf{[MW]:} & 374 & 748 & 1496 & 2244 \\ 
\hline
How many power plants of the size of a single reactor at Fukushima would be needed to generate the same amount of power? \textbf{[COUNT]} & 1,7 & 3,4 & 6,7 & 10,1 \\
\hline
    \end{tabular}
    }\newline 
\end{center}
\hspace{2.5mm}During the second quarter of 2022, the average global user spent more than half of their daily online time on their mobile phone. The average daily consumption was 3 hours and 41 minutes~\cite{kempGlobalStateDigital2022}. Considering in addition the fact that there are 5.48 billion internet users worldwide, we require a total of round about 7480 MW of power for web browsing on a global scale. 

\hspace{2.5mm} If we could reduce 5\% of energy usage consumed by smartphones, we would save 374 Wh. To put this in perspective, this is equivalent to being able to shut down more one Fukushima power reactor.

\hspace{2.5mm} It is worth noting that these estimates are rough. Still, they demonstrate the potential for energy-saving techniques to extend battery life, enhance the user experience, and reduce the environmental impact by conserving energy. The detailed calculations, assumptions, and data can be accessed online \cite{dornauerEnergySavingStrategiesMobile2023}. 

\vspace{2mm}
\end{tcolorbox}

The primary objective of this \ac{slr} is to compile and analyse a comprehensive overview of energy-saving efforts and techniques tested on mobile web applications over the past decade. In particular, our focus is on the impact of mobile web app usage on battery consumption. But we do not exclude other factors that may influence battery life, such as CPU  and network bandwidth utilization.
\\
\\
\\
\\
The specific contributions of this study are as follows:

\begin{enumerate}
    \item  Identifying the key parameters that contribute to energy drain in smartphones due to web app usage. (\hyperref[rq1]{RQ1})
    \item Summarizing the testing methodologies employed in energy-saving experiments conducted on mobile web apps.  (\hyperref[rq2]{RQ2}, \hyperref[rq3]{RQ3})
    \item Analysing and synthesizing the various energy-saving approaches employed in research. (\hyperref[rq4]{RQ4})
\end{enumerate} 

After providing the necessary context and motivation for this \ac{slr}, the subsequent sections of this paper are organized as follows. Subsequently, we discuss in Section \ref{sec:relatedWork} the findings of related secondary studies and emphasize the differentiation from our work. In Section \ref{sec:methodology}, we outline the methodology employed in conducting the \ac{slr}, including the specific research questions. These questions are subsequently addressed in Section \ref{sec:findings}, where we present the findings of the review, including a synthesis of all relevant studies. The implications of these findings are then discussed in Section \ref{sec:discussion}. In Section \ref{sec:threatsofValidity}, we address the potential threats to validity that may have influenced the results of the review and conclude with Section \ref{sec:conclusion}.
\section{Related Work}
\label{sec:relatedWork}
Several primary studies have analysed specific aspects of energy-saving experiments for mobile (web) apps. These research efforts have focused on various topics, such as CPU scheduling (e.g.,~\cite{zhuEventbasedSchedulingEnergyefficient2015}), display interface (e.g.,~\cite{renCamelSmartAdaptive2020}), and offloading (e.g.,~\cite{jeongDynamicOffloadingWeb2020}). The findings of these studies are given in this \ac{slr} presented in Section \ref{sec:findings}. Only secondary studies related to our research are discussed further.

A part of our research is to examine the methodology used for conducting energy-saving experiments. One of the most relevant previous works on this topic was conducted by Munk et al.~\cite{demunkStateArtMeasurementbased2022}. Their literature review of measurement-based experiments on the mobile web, focused on the technical aspects of experimental execution. In distinction, our study places a greater emphasis on energy-related experiments. This can be highlighted by the disparity in the number of studies that have employed the Monsoon power monitor. While Munk et al. reference four studies, our analysis identified 12 papers using this measurement tool. 

A particular objective of our systematic literature review is to find energy-intensive aspects of mobile web applications and to gather information on potential energy-saving strategies. Saving energy can also be considered in a broader context, such as that of green Internet of Things, as demonstrated by Arshad et al. ~\cite{arshadGreenIoTInvestigation2017}. Arshad et al. outlined multiple challenges and proposed a taxonomy of green \ac{iot} techniques. This categorization was based on the technologies which were used in various energy-saving IoT models. Alongside, the authors evaluated different solutions that can be used to reduce the energy consumption of \ac{iot} devices. Overall, the paper provides a comprehensive overview of the different opportunities for energy savings in \ac{iot} but has no explicit implication on how to save software-based energy on mobile applications. 

In contrast to the previously mentioned study, Cruz and Abreu~\cite{cruzCatalogEnergyPatterns2019}, and Meneses-Viveros et al.~\cite{meneses-viverosEnergySavingStrategies2018}  propose strategies that have direct implications for mobile application developers to utilize the system and optimize battery usage efficiently. 

Specifically, Cruz and Abreu present a comprehensive catalogue of energy patterns that can be applied to mobile applications to enhance energy efficiency. This catalogue, comprising 22 design patterns, was developed through a review of commits, issues, and pull requests of over 1700 mobile app repositories and related research. The authors assert that some of these patterns may also apply to the \ac{iot} domain. 
 
Meneses-Viveros et al. focus on the strategic design decisions that mobile app developers can make to conserve energy. Through a \ac{slr}, various strategies for energy saving were identified, with a particular emphasis on those related to the development of applications. These strategies include using mobile computation offloading, prioritizing sequential programming over multithreading and parallel execution, and ensuring proper design and implementation of the graphical user interface.

A direct comparison with our evaluated energy-saving strategies revealed that some of these patterns and strategies are also relevant in the context of mobile web applications. As far as we are aware, the fundamental distinction between our study and prior research pertains to the primary focus of our study on web applications, as opposed to native applications. This serves to emphasize the significance and novelty of our research.

\section{Methodology}
\label{sec:methodology}
\begin{figure}[htbp]
    \centering
    \includegraphics[scale=1]{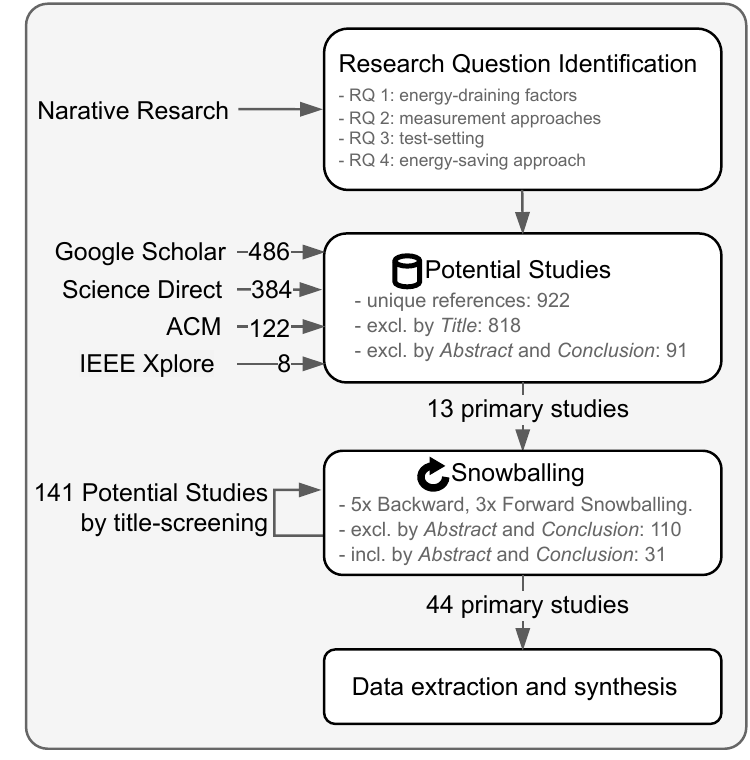}
    \caption{Overview of the study design.}
    \label{fig:overviewStudyDesign}
\end{figure}

In order to conduct this \ac{slr}, we followed Kitchenham's and Charters’s guideline in software engineering~\cite{kitchenhamGuidelinesPerformingSystematic2007} and Wohlin's guidelines for snowballing ~\cite{wohlinGuidelinesSnowballingSystematic2014}. Their instructions are used to identify and synthesize the current state of knowledge on energy-efficiency in mobile web applications and to provide guidance for future research. The systematic research process is introduced in the subsequent subsections. The specific search procedure can be found in Figure \ref{fig:overviewStudyDesign}.

\subsection{Research Questions}\label{sec:researchQuestions}
The objective of research questions \hyperref[rq1]{RQ1}, \hyperref[rq2]{RQ2}, \hyperref[rq3]{RQ3} and \hyperref[rq4]{RQ4}, derived from a narrative research, is to understand the energy-draining aspects of mobile web apps and to aggregate energy-saving approaches from previous research. Additionally, the study examines the measurement process used to obtain comprehensible energy results.
\\
\begin{enumerate}[leftmargin=0.9cm]
    \item[\textbf{RQ1}\label{rq1}] \textit{What hardware components in smartphones are most commonly associated with energy drain caused by mobile web apps?}
    
    The smartphone's energy-draining velocity is closely related to utilizing multiple hardware components. Typical components are \ac{led} display, \ac{cpu} or wireless communication module (\ac{wifi}, mobile networks like 2G, 3G or 4G)~\cite{perrucciSurveyEnergyConsumption2011, tawalbehStudyingEnergyConsumption2016}. By understanding the various parameters that influence energy consumption, developers can optimize the design and development of web applications to minimize energy usage.
    \\
    \item[\textbf{RQ2}\label{rq2}] \textit{What approaches are available to measure the energy consumption of a mobile web application?} 
    
    This research question examines the hardware- and software-based energy measurement setups used in mobile testing environments. It aims to illustrate the different approaches and tools used, as well as discuss their respective advantages and drawbacks.
    \\
    \item[\textbf{RQ3}\label{rq3}] \textit{To obtain comprehensible results, what experimental test settings must be considered to measure a mobile web app's energy consumption?} 
    
    To accurately assess the energy demands of mobile web apps, it is essential to consider a range of experimental test settings that can affect the results. These may include cache settings, connection type, and the specific actions performed within the mobile web app. In this study, we seek to identify the critical test settings that should be taken into account when measuring the energy consumption of mobile web apps.
    \\
    \item[\textbf{RQ4\label{rq4}}] \textit{What approaches exist to reduce the energy consumption of mobile web apps?}
    
    Building on the findings from \hyperref[rq1]{RQ1}, we examine different strategies for decreasing the energy consumption of mobile web apps. To identify common themes, we categorize energy-saving approaches and survey the various aspects of research conducted in each category.
    
\end{enumerate}
\subsection{Literature Search}
 To execute sufficient literature research, we selected a hybrid search strategy to improve the comprehensiveness and sensitivity of the search while minimizing the risk of missing relevant studies~\cite{mouraoInvestigatingUseHybrid2017}. The search process started with an electronic research, followed by forward and backward snowballing~\cite{wohlinGuidelinesSnowballingSystematic2014}.
\subsubsection{Initial search}
    To identify an appropriate start set of search results, we selected the four scholarly search engines: \textit{Google Scholar}\footnote{\url{https://scholar.google.com/}}, \textit{Science Direct}\footnote{\url{https://www.sciencedirect.com/}}, \textit{ACM\footnote{\url{https://dl.acm.org/}} (Association for Computing Machinery)}, \textit{IEEE Xplore\footnote{\url{https://ieeexplore.ieee.org}} (Institute of Electrical and Electronics Engineers)}. All four science search engines allow the use of boolean expressions in search terms. The primary search terms were "energy efficiency" and "mobile web app", which were extended to the following search string:   

    \begin{center}\textit{
        (energy OR power) AND (efficiency OR saving) AND ("mobile web app" OR "mobile web application" OR "PWA" OR "progressive web app" OR "mobile website")}
    \end{center}

    The search was not case-sensitive and limited to the time range from \textit{01 January 2012} to \textit{16 November 2022}. Relevant filtering settings provided by the search engines were also selected to limit the results to those that were relevant. The specific filtering settings are represented in the Review Protocol \cite{dornauerEnergySavingStrategiesMobile2023}. 

\subsubsection{Snowballing}
    To complement our primary studies data set, we chose the snowballing approach provided by Wohlin~\cite{wohlinGuidelinesSnowballingSystematic2014}. Their described concept was used to identify and add relevant studies by starting with a small set of initial studies and iteratively adding more studies that are either cited by the initial studies (backward snowballing) or cite the initial studies (forward snowballing). This process allows us to expand the scope of the review and increase the number of studies considered. In our study, we initially used backward snowballing (BS-1, BS-2, BS-3) until no further studies were found through this process. We then switched to forward snowballing (FS-1). After the first four iterations, we conducted both backward and forward snowballing iteratively (BS-4, FS-2; BS-5, FS-3) until there were no more studies to evaluate. 
     
\subsection{Study Selection}
    We used the inclusion and exclusion criteria listed in Tables \ref{tab:inclusionCriteria} and \ref{tab:exclusionCriteria} to decide which papers to include in the review.
    
    \begin{table}[htbp]
\caption{Inclusion Criteria.}
\label{tab:inclusionCriteria}
\centering
\begin{tabular}{p{0.5cm} p{6.8cm}}
\hline
\rowcolor[rgb]{0.957,0.961,0.961} \textbf{ID} & \multicolumn{1}{c}{\textbf{Criteria}} \\ 
\hline
IC1 & The study considers only web apps  (including \ac{pwa}, mobile websites) accessed via mobile devices. \\
IC2 & The energy-savings are accomplished on client-side. \\
IC3 & The paper should describe the methods used to measure energy consumption. \\
\hline
\end{tabular}
\end{table}

\begin{table}[htbp]
\centering
\caption{Exclusion Criteria.}
\label{tab:exclusionCriteria}
\begin{tabular}{ p{0.5cm} p{6.8cm}}
\hline
\rowcolor[rgb]{0.961,0.961,0.961} \textbf{ID} & \multicolumn{1}{c}{\textbf{Criteria}} \\ 
\hline
EC1 & Paper that has not undergone peer review. \\
EC2 & The paper is not available online or is not written in English. \\
EC3 & The quality of the paper is not up to the necessary scientific standards. \\
EC4 & The paper is not a published journal article, conference proceedings, or workshop paper. \\
EC5 & The paper has already been published in another publication. \\
EC6 & Paper is out of scope: (1) the paper is not related to mobile web development, (2) the paper does not involve an energy saving approach, or (3) the power measurement methods described in the paper are not clear. \\
EC7 & Other conducted literature reviews. \\
\hline
\end{tabular}
\end{table}
    
    The first step in the evaluation process was to read the title of each research and determine if it was suitable for the \ac{slr}. If deemed appropriate, the paper was marked as a candidate for further evaluation. If a syntactic duplicate (a paper with the same title or Digital Object Identifier) was found, the paper was excluded.
    Next, we assessed the eligibility of the research by reading the \textit{Abstract} and \textit{Conclusion} of the paper to decide if it should be added to the primary studies' selection. If a research paper met all the inclusion criteria and none of the exclusion criteria, we examined it in the study.

\subsection{Data Extraction}
If a paper was deemed eligible, it was thoroughly reviewed, and relevant data was extracted, organized, and summarized in accordance with the established extraction framework \cite{dornauerEnergySavingStrategiesMobile2023}. The first author was responsible for performing the data extraction.

\section{Findings}
\label{sec:findings}
In this section, we present the results of the \ac{slr} including the general information and addressing the research questions posed at the beginning of the paper (introduced in Section \ref{sec:methodology}). 

\subsection*{General Findings}
\begin{figure}[htbp]
    \centering
        \begin{tikzpicture}[scale = 0.5]
  
    \pie[
        sum = auto, text = legend
        ]{
             3/Computation offloading,
            11/CPU-efficient scheduling,
            3/Human interaction optimization,
            8/(Pre-)filtering of web-content,
            8/Code-related optimization, 
            3/Caching,
            2/Display-aware content adaption,
            6/Other
        },
    ]
    \end{tikzpicture}
    \caption{Energy-saving approaches.}
    \label{fig:energySavingApproachPie}
\end{figure}
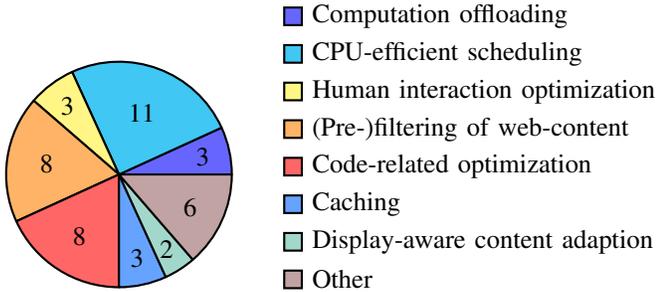

\addcontentsline{toc}{subsection}{\protect\numberline{}General Findings}%
A total of 44 primary studies were included in the review, with five times backward and three times forward snowballing. Thereby, we can't see any significant increase in literature on specific years between 2012 and 2022. 
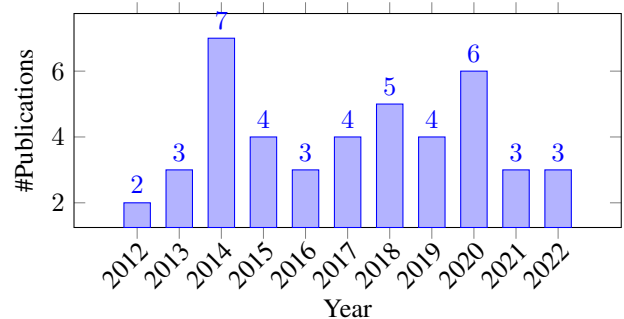
\begin{figure}[htbp]
    \centering
    \begin{tikzpicture}
        \begin{axis}[        
                ybar,   
                width=\linewidth,
                height=0.5\linewidth,
                enlargelimits=0.15,        
                xlabel= Year,
                ylabel= \#Publications,       
                symbolic x coords={2012, 2013, 2014, 2015, 2016, 2017, 2018, 2019, 2020, 2021, 2022},        
                xticklabel style={rotate=45, anchor=north east, inner sep=0pt},
                xtick=data,        
                nodes near coords,      
            ]
            \addplot coordinates {(2012,2) (2013,3) (2014,7) (2015,4) (2016,3) (2017,4) (2018,5) (2019,4) (2020,6) (2021,3) (2022,3)};
        \end{axis}
    \end{tikzpicture}
    \caption{Number of references per year.}
    \label{fig:my_label}
\end{figure}
This suggests that the topic may be receiving less attention than may have been expected with the growth of the World Wide Web~\cite{dachyarKnowledgeGrowthDevelopment2019}. The primary studies included in the review comprised 29 conference papers and 15 journal articles. No workshop paper was identified in this review. The data presented in Figure \ref{fig:energySavingApproachPie} illustrates the distribution of publications on various energy-saving strategies, with a predominance of research focused on CPU-efficient scheduling strategies and code-related optimization.
\subsection*{\textbf{RQ1}: What hardware components in smartphones are most commonly associated with energy drain caused by mobile web apps?}
\addcontentsline{toc}{subsection}{\protect\numberline{}RQ1}%
\begin{table*}[!htbp]
\centering
\caption{Energy-draining hardware influenced by web.}
\label{tab:energyDrainingAspects}
\begin{tabular}{p{2.5cm}|p{3.5cm}p{4.5cm}p{4.7cm}} 
\hline
\rowcolor[rgb]{0.91, 0.91, 0.91} \multicolumn{1}{c}{\textbf{Hardware}} & \multicolumn{1}{c}{\textbf{Display}} & \multicolumn{1}{c}{\textbf{Wi-Fi / GSM module}} & \multicolumn{1}{c}{\textbf{CPU}} \\
\cline{1-4}
{\cellcolor[rgb]{0.91, 0.91, 0.91} 
\textbf{{\newline Energy consuming characteristics}}}& 
\begin{itemize}[leftmargin=3mm]
    \item proportion of light to dark pixels (only relevant for OLED displays)
\end{itemize}
& \begin{itemize}[leftmargin=3mm]
    \item type of connection (\ac{wifi}, 3G, 4G, LTE)
    \item bandwidth of network
    \item network latency 
\end{itemize} &  
\begin{itemize}[leftmargin=3mm]
    \item loading and rendering of web content
    \item human-interaction (repainting)
    \item CPU scheduling 
    \begin{itemize}
        \item frequency scaling
         \item core selection
    \end{itemize}
\end{itemize} \\
\rule[-10pt]{0pt}{2pt}\cellcolor[rgb]{0.91, 0.91, 0.91}\textbf{Count} &
\rule[-10pt]{0pt}{2pt}{\color[rgb]{0.72,0.89,0.90}\rule{3mm}{5pt}}~5  &
\rule[-10pt]{0pt}{2pt}{\color[rgb]{0.72,0.89,0.90}\rule{19mm}{5pt}}~18 &
\rule[-10pt]{0pt}{2pt}{\color[rgb]{0.72,0.89,0.90}\rule{30mm}{5pt}}~34   
\\
\cellcolor[rgb]{0.91, 0.91, 0.91}\textbf{References}
& \hspace{-0.01cm}\cite{renCamelSmartAdaptive2020,dongChameleonColoradaptiveWeb2011,liMakingWebApplications2014,varvelloSheddingLightMobile2021,xuEBrowserMakingHumanMobile2018} 
& \hspace{-0.01cm} \cite{huberComparativeStudyEnergy2022,ayalaEmpiricalStudyPower2017,malavoltaAssessingImpactService2017,duttaCachingReduceMobile2018,qianCharacterizingResourceUsage2014,chowdhuryClientsideEnergyEfficiency2016,alamiEnergyQualityExperience2020,dambrosioEnergyConsumptionPrivacy2016,zhaoEnergyAwareWebBrowsing2013,malavoltaEvaluatingImpactCaching2020,rasmussenGreenMiningEnergy2014,chan-jong-chuInvestigatingCorrelationPerformance2020,dambrosioMobilePhoneBatteries2014,lymberopoulosPocketWebInstantWeb2012,singhRMVRVMParadigmCreating2022,albasirSMoWEnergybandwidthAware2014,heitmannBuildingBetterMobile2020,unluTranscodingWebPages2022}
& \hspace{-0.01cm}\cite{huberComparativeStudyEnergy2022,renAdaptiveWebBrowsing2018,malavoltaAssessingImpactService2017,golestaniCharacterizationUnnecessaryComputations2019,anCommunicatingWebVessels2021,vanhasseltComparingEnergyEfficiency2022,caoDeconstructingEnergyConsumption2017,parkDesignEvaluationMobile2014,shingariDORAOptimizingSmartphone2018,jeongDynamicOffloadingWeb2020,xuEBrowserMakingHumanMobile2018,zhuEventbasedSchedulingEnergyefficient2015,zhuExploitingWebpageCharacteristics2014,rasmussenGreenMiningEnergy2014,zhuGreenWebLanguageExtensions2016,zhuHighperformanceEnergyefficientMobile2013,chan-jong-chuInvestigatingCorrelationPerformance2020,choiOptimizingEnergyEfficiency2019, dambrosioEnergyConsumptionPrivacy2016,fengPESProactiveEvent2019,petersPhaseAwareWebBrowser2018,lymberopoulosPocketWebInstantWeb2012,renProteusNetworkawareWeb2018,iharaRefiningMobileWeb2015,buiRethinkingEnergyperformanceTradeoff2015,singhRMVRVMParadigmCreating2022,shingariDORAOptimizingSmartphone2018,varvelloSheddingLightMobile2021,albasirSmartMobileWeb2013,albasirSMoWEnergybandwidthAware2014,zhuRoleCPUEnergyefficient2015,heitmannBuildingBetterMobile2020,unluTranscodingWebPages2022,yuanUsingMachineLearning2019}\\
\hline
\end{tabular}
\end{table*}
The rate at which a smartphone consumes energy is strongly correlated with the utilization of its various hardware components~\cite{liMeasurementAnalysisEnergy2014}. Table~\ref{tab:energyDrainingAspects} maps the literature that tries to improve this specific hardware's energy-drain and summarizes energy expenditure aspects that are discussed further: 

\textbf{Display:} The display technology of a mobile device significantly influences energy consumption ~\cite{liMeasurementAnalysisEnergy2014}. \ac{lcd} consumes power for backlighting, even when displaying a black screen, while \ac{oled} displays exhibit minimal power consumption when the pixels are turned off. This means that the power consumption of \ac{oled} displays increases as the number of illuminated pixels raise. Changing the visual representation of a mobile web app leads to a specific proportion of light to dark pixels, influencing the energy factor on \ac{oled} displays~\cite{dongChameleonColoradaptiveWeb2011,liMakingWebApplications2014}.  

\textbf{Wi-Fi or GSM module:}
Mobile devices can select between Wi-Fi and cellular (2G, 3G, 4G/LTE) networks to connect to the web. In most cases, mobile data consumes more energy than Wi-Fi, as illustrated in various research~\cite{albasirSMoWEnergybandwidthAware2014, renProteusNetworkawareWeb2018, parkDesignEvaluationMobile2014, zhuRoleCPUEnergyefficient2015}). Bandwidth and latency are key factors that influence energy consumption ~\cite{albasirSMoWEnergybandwidthAware2014}. Bandwidth, typically measured in bytes, refers to the maximum amount of data that can be transferred over a network, while latency, typically measured as round-trip time, refers to the amount of time it takes to transfer data. The energy consumption for transmitting, receiving, and processing data increases with the amount of data transmitted in a specific period~\cite{qianCharacterizingResourceUsage2014,parkDesignEvaluationMobile2014}. However, it is important to note that the time duration of the data transmission also plays a crucial role in determining the overall energy-efficiency. Even though a shorter time frame for data transmission may result in higher energy consumption, it may lead to better energy-efficiency in the long run, as it takes shorter time to load the content~\cite{zhuRoleCPUEnergyefficient2015}.  

\textbf{CPU Processing:}
The energy consumption of a CPU during the loading phase of a mobile web app can be significant as it processes and renders the web content~\cite{caoDeconstructingEnergyConsumption2017}. However, this consumption does not cease after the initial load, as user interactions with the app can also trigger the need for repainting~\cite{xuEBrowserMakingHumanMobile2018, jeongDynamicOffloadingWeb2020,choiOptimizingEnergyEfficiency2019}, thereby increasing CPU utilization. The energy demand of a CPU is determined by two main factors: frequency scaling~\cite{shingariDORAOptimizingSmartphone2018, zhuExploitingWebpageCharacteristics2014,renAdaptiveWebBrowsing2018,zhuHighperformanceEnergyefficientMobile2013} and core selection ~\cite{renAdaptiveWebBrowsing2018}. Frequency scaling refers to the ability of a CPU to operate at different frequencies, depending on the workload. When a CPU runs at a lower frequency, it consumes less energy but at the expense of reduced performance and potentially extended page loading phases. Core selection, on the other hand, refers to the number of cores that are active and running at any given time. Many modern processors have multiple cores, some of which are energy-efficient while  others are high-performance (e.g., \textit{ARM big.LITTLE}). By selectively disabling specific cores, energy consumption can be reduced.
\subsection*{\textbf{RQ2}: What approaches are available to measure the energy consumption of a mobile web application?}\label{sec:RQ2}
\addcontentsline{toc}{subsection}{\protect\numberline{}RQ2}%
\begin{table}[htbp]
\caption{Energy measurement approaches.}
\label{tab:energyMeasurementApproaches}
\begin{tabular}{llm{4cm}}
\hline
\rowcolor[rgb]{0.91, 0.91, 0.91}\multicolumn{1}{c}{\textbf{Approach}} & \multicolumn{1}{c}{\textbf{Count}} & \multicolumn{1}{c}{\textbf{References}} \\ \hline
Software-based & {\color[rgb]{0.72,0.89,0.90}\rule{5mm}{5pt}}~16 &\hspace{-0.01cm}\cite{huberComparativeStudyEnergy2022,singhRMVRVMParadigmCreating2022,petersPhaseAwareWebBrowser2018,malavoltaEvaluatingImpactCaching2020,chan-jong-chuInvestigatingCorrelationPerformance2020,malavoltaAssessingImpactService2017,qianCharacterizingResourceUsage2014,anCommunicatingWebVessels2021,liMakingWebApplications2014,duttaCachingReduceMobile2018,dongChameleonColoradaptiveWeb2011,unluTranscodingWebPages2022,renAdaptiveWebBrowsing2018,jeongDynamicOffloadingWeb2020,ayalaEmpiricalStudyPower2017,vanhasseltComparingEnergyEfficiency2022,ayalaEmpiricalStudyPower2017} \\
Hardware-based & {\color[rgb]{0.72,0.89,0.90}\rule{9mm}{5pt}}~24 & \hspace{-0.01cm}\cite{zhuEventbasedSchedulingEnergyefficient2015, fengPESProactiveEvent2019,dambrosioEnergyConsumptionPrivacy2016,parkDesignEvaluationMobile2014,chowdhuryClientsideEnergyEfficiency2016,buiRethinkingEnergyperformanceTradeoff2015,caoDeconstructingEnergyConsumption2017, zhuRoleCPUEnergyefficient2015,zhuExploitingWebpageCharacteristics2014,shingariDORAOptimizingSmartphone2018,zhuGreenWebLanguageExtensions2016,zhaoEnergyAwareWebBrowsing2013, rasmussenGreenMiningEnergy2014,zhuHighperformanceEnergyefficientMobile2013,albasirSmartMobileWeb2013,dambrosioMobilePhoneBatteries2014,albasirSMoWEnergybandwidthAware2014,xuEBrowserMakingHumanMobile2018,renCamelSmartAdaptive2020,huEnergyOptimizationTraffic2014,heitmannBuildingBetterMobile2020,renProteusNetworkawareWeb2018,yuanUsingMachineLearning2019,varvelloSheddingLightMobile2021}
\\
Estimation-based & {\color[rgb]{0.72,0.89,0.90}\rule{2mm}{5pt}}~4 & \hspace{-0.01cm}\cite{lymberopoulosPocketWebInstantWeb2012,alamiEnergyQualityExperience2020,golestaniCharacterizationUnnecessaryComputations2019,iharaRefiningMobileWeb2015} \\ \hline
\end{tabular}
\end{table}
For the energy measurement, we identified three general approaches: \textit{hardware-based}, \textit{software-based} and \textit{estimation-based}.

One way to measure the energy consumption of mobile web apps is to use \textbf{hardware-based} energy measurement tools. All the provided hardware can be in general described as some kind of multimeter, often able to provide fine-granular data~\cite{zhuEventbasedSchedulingEnergyefficient2015,petersPhaseAwareWebBrowser2018,fengPESProactiveEvent2019,zhuExploitingWebpageCharacteristics2014,zhuHighperformanceEnergyefficientMobile2013} on total device energy consumption. The multimeters are connected to a power link  transmitting energy from a battery or a constant laboratory power supply to the mobile phone. We identified \textit{Monsoon Solutions's power monitor} as the most commonly used hardware \cite{parkDesignEvaluationMobile2014,liMakingWebApplications2014,buiRethinkingEnergyperformanceTradeoff2015,caoDeconstructingEnergyConsumption2017,zhuRoleCPUEnergyefficient2015,albasirSmartMobileWeb2013,albasirSMoWEnergybandwidthAware2014,jeongDynamicOffloadingWeb2020,xuEBrowserMakingHumanMobile2018,renCamelSmartAdaptive2020,choiOptimizingEnergyEfficiency2019,varvelloSheddingLightMobile2021}. We assume that Monsoon's power monitor is often used as it can take up to $5000$ samples per second, and has a  +/- $50 \mu A$ accuracy~\cite{MonsoonPowerMonitor}. 

Furthermore, \textit{ODroid-XU3} is a common tool for energy measurement, equipped with an onboard energy sensor. It is a single-board computer that runs Android or Ubuntu Linux and has a Samsung Exynos 5422 Cortex-A15 core. Its smartphone-like design has made it a useful tool in various energy measurement studies~\cite{zhuEventbasedSchedulingEnergyefficient2015,fengPESProactiveEvent2019,zhuExploitingWebpageCharacteristics2014,shingariDORAOptimizingSmartphone2018,dongChameleonColoradaptiveWeb2011,zhuHighperformanceEnergyefficientMobile2013}. In addition to the previously mentioned approaches, solutions from the brands National Instruments \cite{zhuEventbasedSchedulingEnergyefficient2015, zhuExploitingWebpageCharacteristics2014, zhuHighperformanceEnergyefficientMobile2013}, Agilent \cite{dambrosioEnergyConsumptionPrivacy2016,zhaoEnergyawareWebBrowsing2015,shingariDORAOptimizingSmartphone2018,zhaoEnergyAwareWebBrowsing2013,dambrosioMobilePhoneBatteries2014} and Keithley \cite{dambrosioEnergyConsumptionPrivacy2016,dambrosioMobilePhoneBatteries2014} were also used for data acquisition. One of the drawbacks of hardware-based tools could be their potential price. For instance, the acquisition of the Monsoon's power monitor costs \$989 (January 2023)~\cite{MonsoonPowerMonitor}. 

Therefore, \textbf{software-based} solutions could be profitable. They often enable more fine-grained measurement at the system or code level. In three cases, the \textit{Batterystats tool} was used. Batterystats collects battery data (based on Smartphone supplier profiles) on the device and can be accessed via the Android Debug Bridge, for further analysis, even on a single hardware unit~\cite{huberComparativeStudyEnergy2022, duttaCachingReduceMobile2018,malavoltaAssessingImpactService2017}. An alternative for devices with Qualcomm chips was Trepn Profiler~\cite{ayalaEmpiricalStudyPower2017,malavoltaAssessingImpactService2017,anCommunicatingWebVessels2021,vanhasseltComparingEnergyEfficiency2022}. It was also used by Ahmad et al.~\cite{ahmadReviewMobileApplication2015} with the PowerTutor app to evaluate the performance of dynamic analysis-based energy estimation. Being updated in 2016 the last time, the support for a newer generation of Android is limited,  which could be a drawback for future experiments~\cite{anCommunicatingWebVessels2021}. Another only once used tool~\cite{chan-jong-chuInvestigatingCorrelationPerformance2020} is the Greenspector, a tool (provider) that offers to execute energy measurements based on runs on actual smartphones. A very “rough” approach was executed by Singh~\cite{singhRMVRVMParadigmCreating2022}, who measured the duration between individual battery-\% drops.

The PowerTutor application, which has already been introduced, is an \textbf{estimation model} for energy consumption. Depending upon the model employed, various factors, such as load time or data transmission, may be considered to approximate the energy consumption of a mobile web application. A recent example of energy consumption estimation models was developed by Alami et al.~\cite{alamiEnergyQualityExperience2020}. Those researchers estimated energy consumption based on the battery consumption profile, the current network speed, the sizes of the different versions of the webpage, and the estimated energy consumption of the various versions. Another example of energy estimation models is one presented by Ihara et al.~\cite{iharaRefiningMobileWeb2015} in their research, which utilizes hardware measurements as the basis for their analysis.

\subsection*{\textbf{RQ3}: To obtain comprehensible results, what experimental test settings must be considered to measure a mobile web app’s energy consumption? }
\addcontentsline{toc}{subsection}{\protect\numberline{}RQ3}%
To perform an energy measurement, the mobile device is typically set up in a controlled test environment in which the conditions are held as good as possible constant, and only a few parameters, like connection parameters (e.g.,~\cite{ayalaEmpiricalStudyPower2017, caoDeconstructingEnergyConsumption2017,parkDesignEvaluationMobile2014}) or cache type (e.g.,~\cite{duttaCachingReduceMobile2018,chan-jong-chuInvestigatingCorrelationPerformance2020,malavoltaEvaluatingImpactCaching2020}), are adopted. In the analysed dataset, we have identified common principles that we want to discuss further to obtain  comprehensible test results. 

\textbf{Measurement tool:}
As a first step, it is often recommended to use a power supply that can provide a stable and controllable source of power. This can be achieved using a laboratory power supply holding power on constant level. Also, it is important to choose the correct energy measurement tool in accordance with one's specific requirements. The primary considerations in making this decision include the level of accuracy and the frequency of measurement, as illustrated in \hyperref[sec:RQ2]{RQ2}. 

\textbf{The mobile device:} 
In past experiments, Android devices have been commonly used for measurements, with Samsung~\cite{zhuEventbasedSchedulingEnergyefficient2015,fengPESProactiveEvent2019, qianCharacterizingResourceUsage2014, buiRethinkingEnergyperformanceTradeoff2015, caoDeconstructingEnergyConsumption2017, lymberopoulosPocketWebInstantWeb2012, rasmussenGreenMiningEnergy2014, alamiEnergyQualityExperience2020, unluTranscodingWebPages2022,xuEBrowserMakingHumanMobile2018,choiOptimizingEnergyEfficiency2019,varvelloSheddingLightMobile2021} and  Nexus~\cite{malavoltaAssessingImpactService2017,dambrosioEnergyConsumptionPrivacy2016, parkDesignEvaluationMobile2014, chowdhuryClientsideEnergyEfficiency2016, caoDeconstructingEnergyConsumption2017, shingariDORAOptimizingSmartphone2018,dongChameleonColoradaptiveWeb2011,albasirSmartMobileWeb2013,dambrosioMobilePhoneBatteries2014,albasirSMoWEnergybandwidthAware2014,iharaRefiningMobileWeb2015, ayalaEmpiricalStudyPower2017} devices being the most popular ones. The Android versions differed from Android 4 to 11. 
The energy efficiency improvements observed in mobile web applications can vary significantly depending on the smartphone device used. For example, Bui et al.~\cite{buiRethinkingEnergyperformanceTradeoff2015} conducted experiments comparing the energy usage of a low-end Samsung \textit{S5-E} and a high-end \textit{S5-S} device while using design principles for energy-efficient page loading. They found that the low-spec S5-E saw an 11.7\% reduction in energy consumption, while the S5-S reached a 24.4\% decrease. 

\textbf{Smartphone's idle power fluctuation:}
Also, the idle power of the smartphone plays a crucial role. The idle power consumption refers to the amount of power that a device or system consumes when it is turned on (display active, device unlocked) but not actively being used. Here, the goal is to have 
as few fluctuations as possible. To minimize these fluctuations, the following strategies have been suggested by the literature:  
\begin{enumerate}
    \item Set the display brightness to a constant level and deactivate adaptive brightness~\cite{singhRMVRVMParadigmCreating2022}. Dutta et al.~\cite{duttaCachingReduceMobile2018} even suggest turning the display brightness off entirely if it is irrelevant to the experiment. 
    \item During an experimental run, background tasks such as push notifications or software
    updates can increase idle power consumption. To minimize this impact, it is recommended to have only the essential apps installed on the device. In order to accomplish this, Chan-Jong-Chu et al.~\cite{chan-jong-chuInvestigatingCorrelationPerformance2020} started their experiment with a clean installation of the \ac{os}. In addition, they deinstalled all third-party applications. 
    \item Transmitting and receiving data unrelated to the experiment, can contribute to idle power consumption.  To minimize this effect, airplane-mode can be enabled and only the necessary wireless connection type should be activated~\cite{huberComparativeStudyEnergy2022, rasmussenGreenMiningEnergy2014, dongChameleonColoradaptiveWeb2011,chowdhuryClientsideEnergyEfficiency2016}.
\end{enumerate}

\textbf{Web browser settings:}
In order to retrieve web-based information, a variety of web-browser engines were utilized. Amongst them, Chromium and Firefox were the most frequently employed, with 20 and 9 instances respectively. In every iteration of the experiment, it was considered mandatory to thoroughly purge all persistent website data, including cache, cookies, and browsing history, and to ensure that the browser settings were consistent across all trials. Furthermore, it was observed that certain studies disabled the cache totally~\cite{renCamelSmartAdaptive2020, dambrosioEnergyConsumptionPrivacy2016,zhuExploitingWebpageCharacteristics2014,dambrosioMobilePhoneBatteries2014,renProteusNetworkawareWeb2018,buiRethinkingEnergyperformanceTradeoff2015,heitmannBuildingBetterMobile2020}.

\textbf{The web content:}
The selection of mobile web apps needs to be carefully considered. What kind of website should be tested? Is the content optimized for mobile devices?  An often-seen approach for energy-evaluation experiments was the use Alexa's top 100/500/1000 websites~\cite{shingariDORAOptimizingSmartphone2018,zhaoEnergyAwareWebBrowsing2013,zhuGreenWebLanguageExtensions2016,zhuHighperformanceEnergyefficientMobile2013,petersPhaseAwareWebBrowser2018,albasirSMoWEnergybandwidthAware2014,heitmannBuildingBetterMobile2020,unluTranscodingWebPages2022}. Unfortunately, this website retired on May 1, 2022. In this way, Varvello and Livshits~\cite{varvelloSheddingLightMobile2021} suggest using Tranco\footnote{\url{https://tranco-list.eu/}}, which is a research-oriented top-site ranking page. 

\textbf{Number of test runs:}
Energy test runs should be repeated multiple times to identify and correct errors or inconsistencies in the experimental setup or data collection process. For instance, Zhu et al. executed their experiments three times~\cite{zhuEventbasedSchedulingEnergyefficient2015}, Qian et al.~\cite{qianCharacterizingResourceUsage2014} 6 times, Chowdhury et al.~\cite{chowdhuryClientsideEnergyEfficiency2016} and Ayala et al.~\cite{ayalaEmpiricalStudyPower2017} 20 times or Van Hasselt~\cite{vanhasseltComparingEnergyEfficiency2022} even 30 times.  

\subsection*{\textbf{RQ4}: What approaches exist to reduce the energy consumption of mobile web apps?}
\addcontentsline{toc}{subsection}{\protect\numberline{}RQ4}%
Over the past decade, there has been a plethora of strategies explored for energy conservation in mobile web browsing. These efforts have primarily centred on enhancing the energy efficiency of various system components, such as the wireless interface, CPU cores, and display interface (see \hyperref[rq1]{RQ1}). 
As shown by Table \ref{tab:energySavingApproaches}, the approaches can be broadly classified into three main-categories: \textit{saving processor utilization}, \textit{adoption of web content}, and \textit{reduction of web-traffic} and related sub-categories. It should be noted that some sources may fall into multiple categories or may not be easily categorized, and are thus listed under the category \textit{Other}. \\

\begin{table}
\centering
\caption{Categorization of energy-saving approaches.}
\label{tab:energySavingApproaches}
\begin{tabular}{m{0mm}m{3.6cm}m{1.2cm}m{2.0cm}} 
\hline
\rowcolor[rgb]{0.91,0.91,0.91} \multicolumn{2}{c}{\textbf{Energy-Saving Approach}} & \multicolumn{1}{c}{\textbf{Count}} & \multicolumn{1}{c}{\textbf{References}} \\ 
\hline
\multicolumn{2}{l}{\textbf{Saving processor utilization }} &  &  \\
 & CPU-efficient scheduling & {\color[rgb]{0.72,0.89,0.90}\rule{11mm}{5pt}}~11 &\hspace{-0.01cm}\cite{petersPhaseAwareWebBrowser2018, zhuHighperformanceEnergyefficientMobile2013,zhuEventbasedSchedulingEnergyefficient2015,zhuGreenWebLanguageExtensions2016,zhuExploitingWebpageCharacteristics2014,zhuRoleCPUEnergyefficient2015,choiOptimizingEnergyEfficiency2019, renAdaptiveWebBrowsing2018, renProteusNetworkawareWeb2018,shingariDORAOptimizingSmartphone2018, fengPESProactiveEvent2019}  \\
 & Human interaction optimization & {\color[rgb]{0.72,0.89,0.90}\rule{3mm}{5pt}}~3  & \hspace{-0.01cm}\cite{xuEBrowserMakingHumanMobile2018,renCamelSmartAdaptive2020,yuanUsingMachineLearning2019} \\
 & Computation~offloading & {\color[rgb]{0.72,0.89,0.90}\rule{3mm}{5pt}}~3 &  \hspace{-0.01cm}\cite{jeongDynamicOffloadingWeb2020,parkDesignEvaluationMobile2014,anCommunicatingWebVessels2021} \\ 
\hline
\multicolumn{2}{l}{\textbf{Reduction of web-traffic }} &  &  \\
 & Caching & {\color[rgb]{0.72,0.89,0.90}\rule{3mm}{5pt}}~3 & \hspace{-0.01cm}\cite{lymberopoulosPocketWebInstantWeb2012,duttaCachingReduceMobile2018,malavoltaEvaluatingImpactCaching2020}\\
 & (Pre-)filtering of web-content & {\color[rgb]{0.72,0.89,0.90}\rule{8mm}{5pt}}~8 &\hspace{-0.01cm}\cite{rasmussenGreenMiningEnergy2014,albasirSmartMobileWeb2013,dambrosioMobilePhoneBatteries2014,dambrosioEnergyConsumptionPrivacy2016,heitmannBuildingBetterMobile2020,varvelloSheddingLightMobile2021,albasirSMoWEnergybandwidthAware2014,singhRMVRVMParadigmCreating2022}\\
 \hline \\
\multicolumn{2}{l}{\textbf{Adoption of web content}} &  &  \\
 & Code-related optimization & {\color[rgb]{0.72,0.89,0.90}\rule{8mm}{5pt}}~8 & \hspace{-0.01cm}\cite{iharaRefiningMobileWeb2015, golestaniCharacterizationUnnecessaryComputations2019, buiRethinkingEnergyperformanceTradeoff2015,chan-jong-chuInvestigatingCorrelationPerformance2020,unluTranscodingWebPages2022,caoDeconstructingEnergyConsumption2017, vanhasseltComparingEnergyEfficiency2022,alamiEnergyQualityExperience2020}\\
 & Display-aware content adaption & {\color[rgb]{0.72,0.89,0.90}\rule{2mm}{5pt}}~2 & \hspace{-0.01cm}\cite{dongChameleonColoradaptiveWeb2011,liMakingWebApplications2014} \\
 \hline
\multicolumn{2}{l}{\textbf{Other}}  & {\color[rgb]{0.72,0.89,0.90}\rule{6mm}{6pt}}~6& \hspace{-0.01cm}\cite{huberComparativeStudyEnergy2022,malavoltaAssessingImpactService2017,qianCharacterizingResourceUsage2014,chowdhuryClientsideEnergyEfficiency2016,zhaoEnergyAwareWebBrowsing2013,ayalaEmpiricalStudyPower2017}  \\
\hline
\end{tabular}
\end{table}
\hspace{-0.35cm}\textbf{Saving processor utilization: }\\

There has been a significant amount of research focused on developing \textbf{CPU-efficient scheduling} approaches for mobile web applications that aim to minimize the processing requirements of the device. These approaches differ from traditional OS schedulers in that they consider a wider range of factors, including web content-related parameters and the user experience, when determining the most efficient scheduling of tasks \cite{petersPhaseAwareWebBrowser2018}. This enables them to minimize not only energy consumption but also maintain a high level of \ac{qos}. They reduce the energy by assigning tasks to specific cores and lowering the frequency. In 2013, Zhu et al.~\cite{zhuHighperformanceEnergyefficientMobile2013} published a paper in which they described the development of a predictive model for optimizing the scheduling of webpages on core and frequency configurations based on website characteristics. Their model met strict performance requirements
and effectively scheduled webpages for optimal performance. In a follow-up paper, Zhu et al.~\cite{zhuEventbasedSchedulingEnergyefficient2015} further considered the user experience and proposed an event-based scheduling approach that executes events with just enough energy to meet \ac{qos} requirements. A year later, the Zhu et al. proposed \textit{GreenWeb}~\cite{zhuGreenWebLanguageExtensions2016},  a set of language extensions that allow web developers to express \ac{qos} abstractions as program annotations. In consideration of the \ac{qos}, the best minimal scheduling configuration of the asymmetric multiprocessing architecture was chosen. \textit{Webtune} developed by Choi et al.~\cite{choiOptimizingEnergyEfficiency2019} also considers \ac{qos} aspects in their research. During the execution, \textit{Webtune} calculates the optimal speed of the browser's processes using reinforcement learning. In 2018, Ren et al.\cite{renAdaptiveWebBrowsing2018} employed a machine learning approach to predict the optimal processor settings for rendering web content. The approach was based on various factors such as the content of the web page (related to Zhu et al.'s research~\cite{zhuExploitingWebpageCharacteristics2014} on exploiting webpage characteristics), network status (related to Zhu et al.'s research on the role of CPU in energy-efficient web browsing in \cite{zhuRoleCPUEnergyefficient2015}), and optimization goals. Also, DORA \cite{shingariDORAOptimizingSmartphone2018} was introduced during this year that concentrated not only on the web content but also on factors like dynamically-varying architecture (such as core utilization, core temperature, and cache) and system conditions. 

Besides optimizing the loading and rendering phase of the website content, also the user's interaction phase controlling the mobile web application provides energy-saving opportunities, labelled with \textbf{human interaction optimization}. Various papers~\cite{xuEBrowserMakingHumanMobile2018, renCamelSmartAdaptive2020, yuanUsingMachineLearning2019}, investigate the scrolling and pinching phase during user's interaction. Xu et al. developed \textit{eBrowser}, a framework designed to optimize energy efficiency in human-computer exchange. When the interaction event update function is called too frequently, it can place a burden on the device's CPU and increase power consumption. To minimize the frequency of these events, \textit{eBrowser} adjusts the interaction event rate (i.e., the rate at which user event triggers occur on a screen)~\cite{xuEBrowserMakingHumanMobile2018}. Xu et al. did not consider the use of specific heterogeneous multicore systems, while Ren et al.~\cite{renCamelSmartAdaptive2020} as well as Yuan et al. \cite{yuanUsingMachineLearning2019} later addressed this issue in their works. They employed various machine learning models to estimate the minimum number of frames per second that can be achieved for a given user interaction on a web page. The optimal processor settings for heterogeneous multicore hardware were then selected in a way that balanced responsiveness with energy efficiency. Achieving this balance was also taken into account for the Proactive Event Scheduling (PES) approach by Feng and Zhu \cite{fengPESProactiveEvent2019}. The concept behind PES is to predict future events, based on a combination of statistical inference and application code analysis, and subsequently synchronize scheduling decisions throughout all events on a global scale. Using data mining \cite{zhaoEnergyAwareWebBrowsing2013}, was another method to predict the user's reading time on webpages and switch the mobile phone to low power if the reading time exceeded a certain threshold. 

High-intensity computation tasks like 3D rendering, Virtual Reality and AI simulation~\cite{jeongDynamicOffloadingWeb2020} on the web can put a strain on a smartphone's CPU and GPU, causing them to run at their peak performance and drain a lot of energy. \textbf{Offloading}, or outsourcing, these tasks to a server, can help to conserve battery capacity and improve performance. Park et al.~\cite{parkDesignEvaluationMobile2014} proposed an offloading system that used a built-in proxy to categorize \ac{js} code into computing-extensive and lightweight code. The lightweight code was executed on the smartphone, while computationally intensive tasks were executed on the server. However, the effectiveness of offloading is highly dependent on the stability of the connection between the server and the client, which can prevent energy savings on the client side. In response to this issue, An and Tilevich~\cite{anCommunicatingWebVessels2021} introduced the concept of \ac{cwv}. \ac{cwv} constantly monitors the network and determines when \ac{js} code should be offloaded. To facilitate the development of offloading systems, Jeong et al.~\cite{jeongDynamicOffloadingWeb2020} came up with \textit{Snapshots}, which allows saving the offloading state of both DOM and \ac{js} computations at any time. \\

\hspace{-0.35cm}\textbf{Reduction of web-traffic:}\\ 

\textbf{Caching} is a technique to store copies of web pages, images, and other types of content locally on a mobile device so that they can be quickly retrieved the next time the same content is requested. The avoidance of transmitting similar content again can reduce energy consumption. 
Before the emergence of PocketWeb in 2012 by Dimitrios Lymberopoulos et al.~\cite{lymberopoulosPocketWebInstantWeb2012}, caching techniques were primarily energy-efficient when dealing with static web content on mobile devices. With the progression of web content becoming increasingly dynamic and frequently changing, traditional caching methods have become inadequate. To address this challenge, PocketWeb implemented a predictive AI-based page-loading mechanism, which segments web pages into regularly visited and spontaneously visited categories for individual users. This approach allows users to benefit from faster browsing speeds while maintaining or reducing radio energy dissipation. 
As reported by Dutta et al.~\cite{duttaCachingReduceMobile2018}, traditional caching mechanisms have limited applicability in the context of mobile devices due to architectural differences between conventional client-server systems and mobile communication infrastructures. To address this issue, the authors propose two caching approaches: (a) response caching, which involves storing complete HTTP responses at the document level, and (b) object caching, which is more suitable for storing fine-grained data in JSON objects. Recent research by Malavolta et al.~\cite{malavoltaEvaluatingImpactCaching2020} has examined the effects of empty and populated caching on Progressive Web Applications (PWAs). The study found that there is no significant difference in energy consumption between the two caching strategies. 

\textbf{(Pre-)filtering} of web content refers to the process of evaluating and selecting web content before it is presented to the user. This is often seen for ads, preserving consumers' online privacy and security. Besides this benefit, blocking mechanisms and privacy filters reduce energy consumption in most cases. This can be achieved through a combination of bandwidth savings – less network request coming from ads – and reducing the load on the CPU by limiting animations, visualizations, and rendering~\cite{rasmussenGreenMiningEnergy2014,albasirSmartMobileWeb2013,dambrosioMobilePhoneBatteries2014,dambrosioEnergyConsumptionPrivacy2016, qianCharacterizingResourceUsage2014, caoDeconstructingEnergyConsumption2017}. This can lead to significant energy savings on mobile devices, as seen with the emergence of the Brave browser, which blocks ads by default to reduce energy consumption~\cite{heitmannBuildingBetterMobile2020, varvelloSheddingLightMobile2021}. All this is done on the client side. But it must also be considered that ads cannot be blocked entirely, as most web apps rely on them for revenue. To solve the issue, Albasir et al.~\cite{ albasirSMoWEnergybandwidthAware2014} introduced a model that adopts web-content on server-side. Based on the smartphone's current battery level and the network type, the number of ads is adjusted. That means that ads were already removed on the server-side. This preserves bandwidth already before data is transmitted to the client while balancing the satisfaction of advertisers as well as the end users. Likewise, Sing~\cite{singhRMVRVMParadigmCreating2022} adopted the classical Model–View–Viewmodel, which can make code more CPU intensive due to data transformations and filtering required between data models and UI elements. Additionally, an overhead of unused data is often observed. Therefore, they moved both the model and viewmodel of the application to the server-side to alleviate the mobile device's workload.

Cao et al.~\cite{caoDeconstructingEnergyConsumption2017} and Qian et al.~\cite{ qianCharacterizingResourceUsage2014}  not only considered caching and filtering as potential strategies for reducing energy consumption, but also proposed the utilization of varying compression levels as a viable option (e.g. HPACK was used in HTTP2~\cite{chowdhuryClientsideEnergyEfficiency2016}). The utilization of compression techniques can aid in minimizing the transmission time of scripts across networks, thus reducing delay. However, it is important to note that the decompression process may also have an impact on energy consumption, and the level of compression chosen can affect this impact~\cite{caoDeconstructingEnergyConsumption2017}. 

Furthermore, the establishment of a connection between the mobile client and server-side is of importance in regard to energy consumption. Common strategies employed for the transmission of data in real-time applications include polling, long polling, and web sockets. With exception to large messages, where Long Polling should be used, Ayala et al. \cite{ayalaEmpiricalStudyPower2017} claim that Web Sockets are the greenest strategy. 
For \ac{pwa}s, the utilization of Service Workers (proxy middleware) has evolved, providing generic entry points for communication between the web application and the server, such as push notifications and message passing. The usage of Service Workers should not have a significant impact on energy consumption for mobile devices according to \cite{malavoltaAssessingImpactService2017, huberComparativeStudyEnergy2022}.\\

\hspace{-0.35cm}\textbf{Adoption of web content:}\\

In contrast to the other categories, \textbf{code-related optimization} will focus on energy-saving methodologies that directly alter web content, explicitly using \ac{js}, \ac{css}, and \ac{html}. 

An overhead of computations arises if web content has pitfalls in its design. The intelligent organization of web content should be based on the popularity of the content~\cite{iharaRefiningMobileWeb2015}, with the most frequently accessed items at the top and less popular ones below. This will improve the user experience by making it easier to find desired content and minimize power consumption caused by excessive scrolling.  Also, unused \ac{js} and \ac{css} code~\cite{golestaniCharacterizationUnnecessaryComputations2019} or overlapping elements, hiding specific layers~\cite{golestaniCharacterizationUnnecessaryComputations2019, buiRethinkingEnergyperformanceTradeoff2015}, are code-related pitfalls. Therefore, the usage of tools like Lighthouse are essential for mobile web apps, as shown by Chang-Jong-Chu et al.~\cite{chan-jong-chuInvestigatingCorrelationPerformance2020}. The research revealed that a lower performance score (a weighted average score based on different metrics like First Contentful Paint or Time to Interactive) in Lighthouse correlates to higher energy consumption. One of Google's lighthouse suggestions was also discussed by Unlu and Yesilada~\cite{unluTranscodingWebPages2022}: Minifying \ac{js} and \ac{css} content involve the elimination of superfluous characters from the source code of a website, including white space, new line characters, and comments. In this way, fewer Bytes must be transferred to the client. Similarly, the consolidation of \ac{js} and \ac{css} files~\cite{unluTranscodingWebPages2022, caoDeconstructingEnergyConsumption2017}  into a single file serves to decrease the number of Hypertext Transfer Protocol requests.  As an alternative to \ac{js} code that has high-computation requirements, the launch of WebAssembly constitutes a performance equally energy-efficiency enhancement~\cite{vanhasseltComparingEnergyEfficiency2022}. WebAssembly can be integrated as a virtual machine into the \ac{js}-environment. 
The condition of the smartphone may also play a role in the adoption of dynamic content. Factors such as those mentioned by Alami et al.~\cite{alamiEnergyQualityExperience2020} may also be relevant, including the average browsing time, estimated current network speed, and remaining battery capacity. Based on that information, the server provides the most appropriate available version (e.g., versions with different resolutions of images or amount of animations) of a web app that maintains the user's experience by minimizing the device's energy consumption. 

The objective of \textbf{display-aware}  content adaptation is to modify web content in order to reduce power consumption in \ac{oled} displays, often achieved by maximizing dark pixels and minimizing light pixels. In 2012, Dong et al.~\cite{dongChameleonColoradaptiveWeb2011} introduced the Chameleon browser, which renders webpages with power-efficient colour schemes. However, this approach had a limitation of manual adoption of colour schemes. A subsequent solution was proposed by Li et al.~\cite{liMakingWebApplications2014}. They developed an automated method for analysing web content, generating a Colour Conflict Graph and using it to create power-efficient colour themes for websites.
\section{Discussion}
\label{sec:discussion}

This section delves into the key findings that emerged from our study. We provide recommendations for researchers and web developers and divide the discussion into two subsections. The first subsection deals with the insights gleaned from the energy-measurement experiment (\hyperref[rq2]{RQ2}, \hyperref[rq3]{RQ3}), while the second one focuses on the energy-saving approaches (\hyperref[rq1]{RQ1}, \hyperref[rq4]{RQ4}).

\subsection{Insights and recommendations for energy-saving experiments }
Over the past decade, numerous studies have been conducted to investigate energy-saving strategies on mobile devices. The majority of these studies were performed on Android devices, due to the open-source nature of the Android operating system, which facilitates experimentation. However, a notable lack of energy-measurement studies utilizing iOS devices in this field has been observed. The specific energy saving strategies may vary depending on the specific characteristics of the mobile web app and the device it is running on. 

The Monsoon Power Monitor is a recommended device for conducting energy-saving approach because it has been used successfully in previous experiments and provides accurate measurements. Additionally, it is easy to put into operation. Although, cost-effective software-based measurements are popular in energy conservation research, yet they may lack the precision of hardware-based measurements. The software-based energy-profiler found by our review (BatteryStats, Trepn, PowerTutor) have an accuracy between 86\% and 97.7\%, according to Hoque et al.~\cite{hoqueModelingProfilingDebugging2015}. 

Beside the imprecision of measurement tools, multiple other factors also influence the accuracy of energy-measurement. The cumulated findings of \hyperref[rq3]{RQ3} may serve as guideline. The wide variations of different test settings, test devices and measurement devices make it (incredible) difficult to compare improvement factors. Therefore, we renounce to mention concrete values, like percentages or improvement factors, as they are hardly comparable. In this way, the generalizability is often missing. Overall, a broader range of individual devices should be selected by researchers, to illustrate the energy-savings. 

\subsection{Insights and recommendations for energy-saving strategies}
Overall, the studies typically focused on energy consumption as a primary aspect, often in conjunction with other factors such as performance, bandwidth, and adoption of web content considerations. Improving, one of those aspects of mobile web apps, correlates directly with an energy-improvement. 

The implementation of different energy-saving strategies often requires a diverse set of technical skills. These skills include proficiency in hardware, knowledge about operating systems and web browsers, as well as expertise of web content optimization techniques. 

Also, what we have seen, is that it becomes important to consider the trade-offs between energy efficiency and other factors such as performance, usability, and profit when choosing an approach for reducing the energy consumption of mobile web apps. 

Investigating the different energy-saving strategies and comparing it with L. Cruz and R. Abreu.~\cite{cruzCatalogEnergyPatterns2019} mobile  energy patterns, we determine that these are not only useable for native apps, but also provide recommendations for mobile web apps. The review revealed that the patterns Push Over Poll, Reduce Size, \ac{wifi} Over Cellular, Cache or Avoid Extraneous Graphics and Animations are already covered by the reviewed papers. 
\section{Threats to Validity}
\label{sec:threatsofValidity}
Threats to Validity refer to factors that may undermine the reliability and generalizability of a study~\cite{wohlinExperimentationSoftwareEngineering2012}. To recognize potential sources of bias, we checked against a map of prevalent Threats to Validity in software engineering formulated by Zhou et al.~\cite{zhouMapThreatsValidity2016}. 

\textbf{Construct Validity:} To ensure an appropriate retrieval process of relevant papers, we adhered to well-established guidelines. Precisely, procedures from Kitchenham et al.~\cite{kitchenhamGuidelinesPerformingSystematic2007} were followed throughout the three phases \ac{slr}: planning, conducting and reporting the review. We conducted a narrative literature review prior to starting our systematic literature review (SLR) to exclude irrelevant research questions and establish appropriate selection criteria. Additionally, we employed guidelines by Wohlin~\cite{wohlinGuidelinesSnowballingSystematic2014} to specifically utilize forward and backward snowballing as a search method.

\textbf{Internal Validity:} To address internal threats, we established a detailed research design plan (outlined in Section \ref{sec:methodology}). One potential internal threat may be that one author solely conducted the extraction process and synthesis of the paper. To mitigate this risk, we have made our replication package available \cite{dornauerEnergySavingStrategiesMobile2023} for other researchers to independently replicate the search process. In addition, we have implemented sanity checks on the extracted data. Conversely, the selection by only one author mitigates the risk of inconsistencies in the selection of studies and data extraction.

\textbf{External Validity:} Due to the study design, gray literature (e.g.~\cite{northEnergyEfficientWebsitea, lalloueListBadGreen2023}) was not included, which may provide insights into energy-saving strategies. As opposed to our claim on science, this gray literature has not undergone rigorous energy experiments and their potential relevance may be limited, due to incomplete research information. To avoid missing potential scientific papers, we used various research engines and employed backward and forward snowballing to conclude the search process after no further relevant papers were found. We anticipated that a time range of ten years would be sufficient for the review, considering the rapid growth and advancements in mobile technology and internet speed. For instance, 80\% of the published literature in ACM, was published within this ten years time frame. 

\textbf{Conclusion Validity:} The Conclusion Validity of our study was ensured by mitigating potential research biases through consultation with the co-author and other researchers. Furthermore, we have adhered to established guidelines for the literature search and have fully disclosed all findings and materials obtained during the search process.
\section{Conclusion}
\label{sec:conclusion}
The significance of energy conservation is now more apparent than ever, especially in light of recent circumstances ~\cite{GlobalImpactWar2022}. As illustrated by our work, 5\% reduction in energy consumption, for example, could result in an additional 39 minutes of battery life for an average 2022 smartphone. When extrapolated to all mobile devices, this reduction in energy consumption would be sufficient to shut down one of the nuclear reactors in Fukushima. 

Through the review and analysis of 44 scientific sources published between 2012 and 2022, this study provides an overview of the energy-saving strategies of mobile web apps. The identified findings disclosed effective approaches for reducing processor utilization, cutting down on web-traffic, and adopting web content efficiently. Additionally, this \ac{slr} considers commonly used measurement techniques and guidelines for conducting related experiments in the future. Our study is a beneficial resource for those looking for inspiration and guidance in implementing energy-saving strategies for mobile web apps.

\section{Acknowledgment}
This work has been supported by and done in scope of the ITEA3 SmartDelta project, which has been funded by the Austrian Research Promotion Agency (FFG, Grant No. 890417). 
\bibliographystyle{IEEEtran}

\end{document}